\documentstyle[multicol,aps,epsfig]{revtex}
\begin{document}

\draft
\title{Photoemission Quasi-Particle Spectra of Sr$_2$RuO$_4$}
\author{A. Liebsch and A. Liechtenstein}
\address{Institut f\"ur Festk\"orperforschung, Forschungszentrum
J\"ulich, 52425 J\"ulich, Germany}
\date{\today }
\maketitle

\begin{abstract}
Multi-band quasi-particle calculations based on perturbation theory and 
dynamical mean field methods show that the creation of a photoemission hole 
state in Sr$_2$RuO$_4$ is associated with a highly anisotropic self-energy. 
Since the narrow Ru-derived $d_{xz,yz}$ bands are more strongly 
distorted by Coulomb correlations than the wide $d_{xy}$ band,
charge is partially transferred from $d_{xz,yz}$ to $d_{xy}$, thereby  
shifting the $d_{xy}$ van Hove singularity close to the Fermi level.   
\end{abstract}

\pacs{PACS numbers: 79.60.Bm, 73.20.Dx, 74.70.Ad, 74.80.Dm}

\begin{multicols}{2}

Angle-resolved photoemission spectroscopy is one of the key techniques
providing detailed information on the Fermi surface topology of high
temperature superconductors. In the layered copper oxide compounds   
(cuprates) high-resolution photoemission spectra can be obtained below
and above the superconducting transition temperature and  for different 
hole doping regimes. Although de Haas--van Alphen (dHvA) experiments 
in principle yield more reliable bulk Fermi surface data, they are less 
useful for the investigation of cuprates since they require extremely  
pure samples. Thus, it has so far not been possible for any of the 
high $T_c$ cuprates to obtain consistent Fermi surface data from both 
photoemission and dHvA measurements.

The detection of superconductivity in Sr$_2$RuO$_4$ \cite{maeno} is of great 
importance since this system is the only layered perovskite compound known 
so far that is superconducting in the absence of copper and without requiring 
doping. Thus, a critical comparison of photoemission Fermi surface 
data with those derived from dHvA measurements is feasable. Surprisingly, 
independent studies of the dHvA effect \cite{mackenzie} and angle-resolved 
photoemission \cite{yokoya,lu,puchkov} yield highly contradictory Fermi 
surface topologies. This discrepancy raises serious questions concerning the 
interpretation of photoemission data also in cuprate superconductors.
         
Because of the layered structure of Sr$_2$RuO$_4$, the electronic bands
close to the Fermi level may be qualitatively understood in terms of a
simple tight-binding picture. These bands are derived mainly from  
Ru $t_{2g}$ states. The wide $xy$ band exhibits two-dimensional character, 
while the narrow $xz$ and $yz$ bands are nearly one-dimensional. 
All three bands are roughly 2/3 occupied, giving about 4 Ru $d$ electrons 
per formula unit. Density functional calculations based on the local 
density approximation (LDA) \cite{oguchi,mazin} place the $(\pi,0),(0,\pi)$ 
saddle point van Hove singularity of the $xy$ band about 60 meV {\it above} the 
Fermi energy.
Taking into account gradient corrections slightly lowers this singularity
to about 50 meV above $E_F$ \cite{mazin}. Fig.~1 provides a qualitative 
picture of the $t_{2g}$ bands and of the Fermi surface exhibiting one 
hole sheet ($\alpha$) and two electron sheets ($\beta$, $\gamma$). 
Whereas the dHvA data \cite{mackenzie} are consistent with these results,
photoemission spectra reveal a fundamentally different topology
\cite{yokoya,lu,puchkov}: the $xy$ van Hove singularity near $M$ appears 
{\it below} the Fermi level, converting the $\gamma$ sheet from 
electron-like to hole-like. Nevertheless, both experiments are reported to 
be in accord with Luttinger's theorem. 

%1
\begin{figure}[b!] 
\begin{center}     
\vskip -0.7cm 
\epsfig{figure=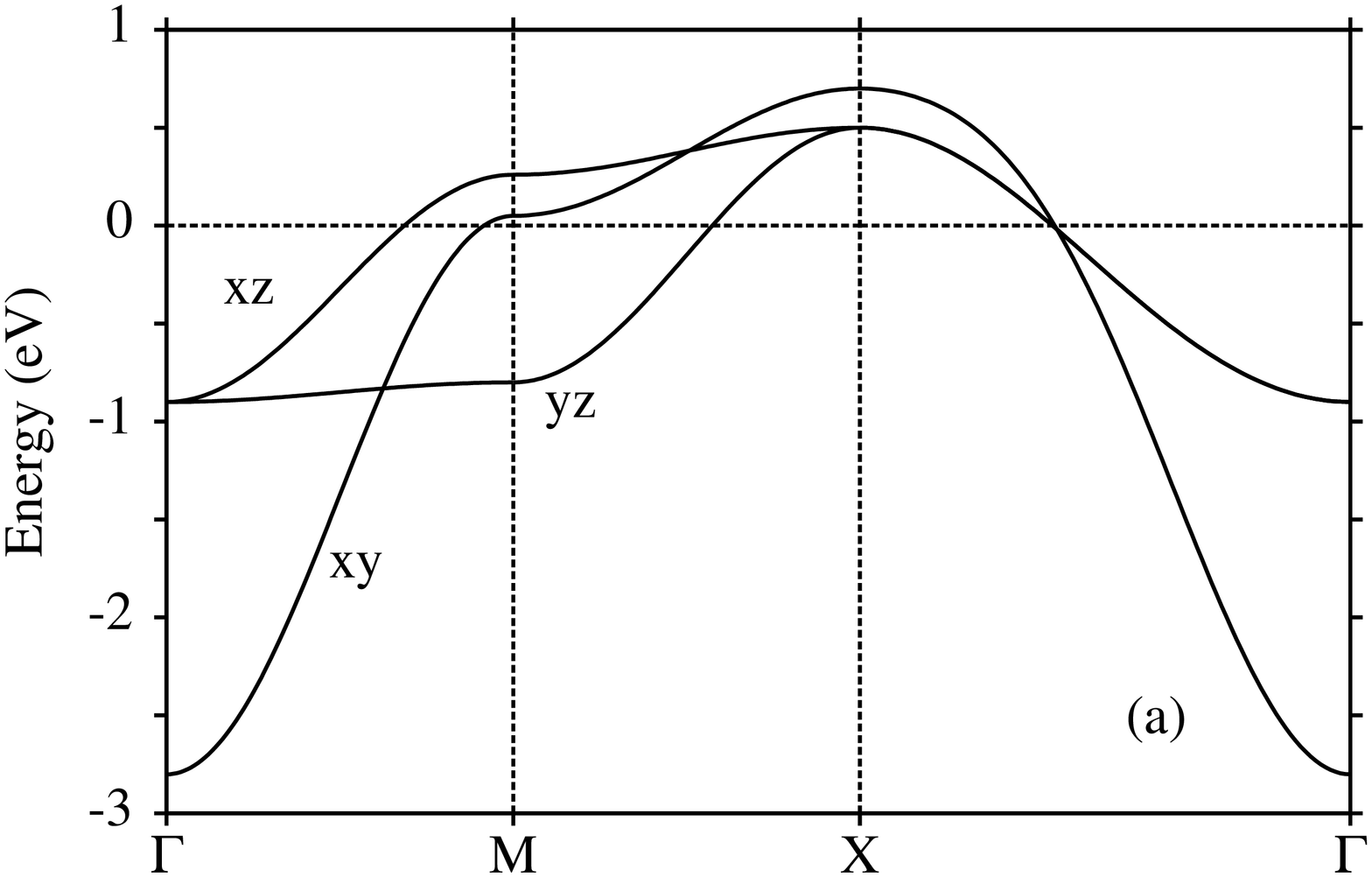,width=5cm,height=7cm,angle=-90}
\vskip -0.3cm
\end{center}
\begin{center}     
\epsfig{figure=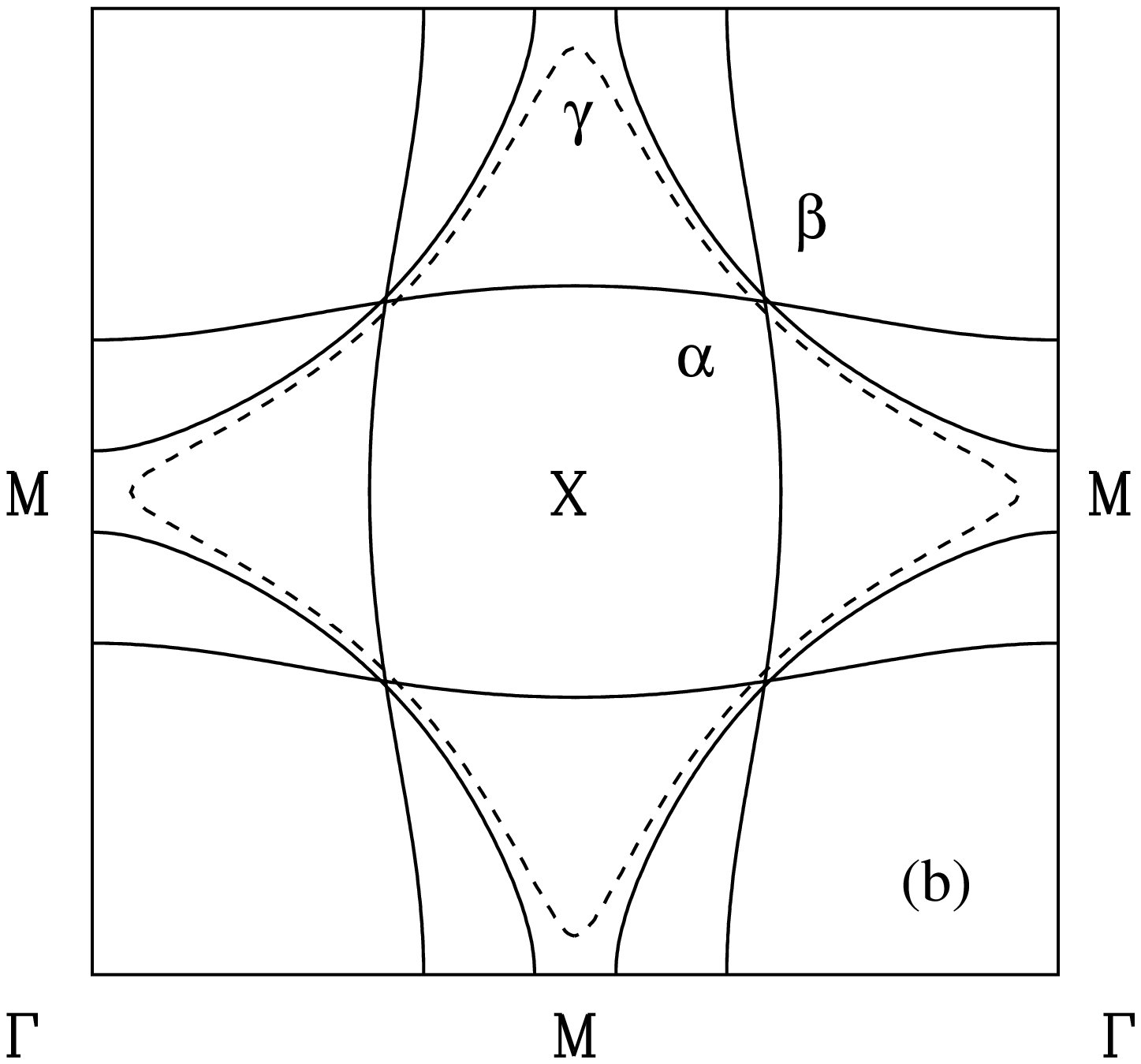,width=5cm,height=5cm,angle=0}
\vskip -0.1cm
\end{center}
\caption{ 
(a) Dispersion of $t_{2g}$ bands of Sr$_2$RuO$_4$ in simplified
two-dimensional Brillouin Zone ($E_F=0$).
(b) Solid lines: Fermi surface consistent with LDA band 
structure and dHvA measurements, with hole sheet $\alpha$ and electron 
sheets $\beta$, $\gamma$ (after accounting for hybridization). 
Dashed line: approximate $xy$ Fermi surface derived from photoemission, 
indicating that $\gamma$ is hole-like. 
}\end{figure}

Various effects have been proposed to explain this discrepancy between
dHvA and photoemission data. (a) Since photoemission is surface sensitive, 
a surface-induced reconstruction associated with the breaking
of bonds could lead to a modification of Ru--O hopping and a shift of 
$t_{2g}$ bands.
Estimates based on bulk calculations indicate that this 
mechanism could push the $xy$ van Hove singularity below $E_F$ \cite{mazin}. 
Actual surface electronic structure calculations for Sr$_2$RuO$_4$ \cite{fang}, 
however, show that this singularity remains above $E_F$ even in the presence 
of surface relaxation. (b) Slightly different
doping levels, temperatures, and magnetic fields used in both experiments
could result in different Fermi surfaces, but these effects are 
believed to be too small to explain the main discrepancy \cite{mac98}.
(c) That the interpretation of the dHvA data is incorrect seems unlikely
in view of the large body of experience available for this method.
(d) The interpretation of the photoemission spectra, on the other hand, 
is non-trivial because of the quasi-particle character of the hole state 
created in the experiment. This aspect of the data has been ignored so far 
and is the focus of the present work.

According to the reduced dimensionality of Sr$_2$RuO$_4$, creation of a   
photohole should be associated with highly anisotropic screening  
processes which reflect the nature of the different electronic states
involved. As shown in Fig.~1, the relevant bands near $E_F$ comprise a 
roughly 3.5~eV wide band formed by in-plane hopping between Ru $d_{xy}$ 
and O $2p$ orbitals, and $1.4$~eV narrow $d_{xz}$, $d_{yz}$ bands. 
Assuming an on-site Ru $dd$ Coulomb interaction $U\approx1.5$~eV, we have 
the intriguing situation: $W_{xz,yz}<U<W_{xy}$, where $W_i$ is the width 
of the $i^{th}$ \,$t_{2g}$ band. A value $U\approx1.5$~eV was in fact 
deduced from the observation of a valence band satellite in resonant 
photoemission from Sr$_2$RuO$_4$ \cite{satellite}. According to this 
picture, intra-atomic correlations have a much larger effect on the 
$xz,yz$ bands than on the $xy$ band, giving rise to a strongly anisotropic 
self-energy. Because of the $\sim2/3$ filling of the $xz,yz$ bands, their 
narrowing, combined with Luttinger's theorem, leads to a charge flow from 
the $xz,yz$ bands to the $xy$ band. As we discuss below, for reasonable 
values of $U$ this charge transfer is large enough to push the $xy$ 
van Hove singularity close to or even below the Fermi level.
    
Since we are concerned with the qualitative influence of multi-band 
correlations on quasi-particle spectra, we consider for simplicity    
next-nearest-neighbor tight-binding bands of the form 
\,$\varepsilon(k)=-\varepsilon_0 - 2 t_x\cos ak_x - 2 t_y\cos ak_y
                                + 4 t'\cos ak_x\cos ak_y$, 
where 
\,$(\varepsilon_0, t_x, t_y, t')=$     
   ($0.50$, $0.44$, $0.44$, $-0.14$), 
   ($0.24$, $0.31$, $0.045$, $0.01$),
   ($0.24$, $0.045$, $0.31$, $0.01$) eV for $xy,xz,yz$, respectively       
(see Fig.~1). These parameters ensure that the $xy$ band has edges  at   
$-2.8$ and 0.7~eV, with a van Hove singularity at 0.05~eV, and  
the $xz,yz$ bands have edges  at $-0.9$ and 0.5~eV, with van Hove 
singularities at $-0.80$ and 0.26~eV, in agreement with 
the LDA band structure \cite{mazin}. 

Next we specify the on-site Coulomb and exchange integrals which we
use in the self-energy calculations discussed below. In the present
case involving only $t_{2g}$ states, there are three independent elements
($i\ne j$) \cite{kanamori}:
\,$U =\langle ii\vert\vert ii\rangle$,     
\,$U'=\langle ij\vert\vert ij\rangle$, and  
\,$J =\langle ij\vert\vert ji\rangle 
     =\langle ii\vert\vert jj\rangle
     = (U-U')/2$, 
where $i=1\ldots3$ denotes $xy,xz,yz$.
Thus, the Hartree-Fock energies are 
\,$\Sigma_1^{HF}=n_1 U + 2 n_2 (2U'-J)$\, and  
\,$\Sigma_{2,3}^{HF}=n_1 (2U'-J) + n_2 (U + 2U'-J)$.
As the band occupations $n_i$ are rather similar, it is convenient 
to define the average occupation $\bar n$, so that 
$n_1=\bar n - 2\delta$, \,$n_{2,3}=\bar n +\delta$, % with $\delta=0.01$.
and       
\,$\Sigma_1^{HF}=5\bar n(U-2J) + 2\delta(U-5J)$,  
\,$\Sigma_{2,3}^{HF}=5\bar n(U-2J) -  \delta(U-5J)$. 

It is instructive to consider the second-order contribution to the local
self-energy since the key point, namely, the large difference between 
the quasi-particle shifts of the $xy$ and $xz,yz$ bands, 
can already be illustrated in this approximation. Because the $t_{2g}$
bands do not hybridize, the self-energy has no off-diagonal elements.
The imaginary parts of the diagonal second-order Coulomb and exchange 
terms are given by
\begin{equation}
 {\rm Im}\,\Sigma_i(\omega) = \pi\!\sum_{jkl}\! R_{jkl}(\omega)\,
      \langle ij\vert\vert kl\rangle   \,\big[ 
    2 \langle kl\vert\vert ij\rangle -  
      \langle kl\vert\vert ji\rangle    \big]  \label{sigma}
\end{equation}                                                          
where
\begin{eqnarray} 
 R_{jkl}(\omega)&=& \Big(\int_0^{ \infty} \!\!\!
                         \int^0_{-\infty}\int^0_{-\infty} \!\! + 
                         \int^0_{-\infty}  
                         \int_0^{ \infty}\!\!\!\int_0^{ \infty} \! \Big) \,
                 d\omega_1d\omega_2d\omega_3               \nonumber \\
    && \!\!\times \,\rho_j(\omega_1)\rho_k(\omega_2)\rho_l(\omega_3)
                  \,\delta(\omega+\omega_1-\omega_2-\omega_3).
\end{eqnarray} 
Here, $\rho_j(\omega)$ denotes the single-particle density of $t_{2g}$ states. 
Exploiting the symmetry properties of the Coulomb matrix elements, 
Eq.~(\ref{sigma}) reduces to
\begin{eqnarray}
   {\rm Im}\,\Sigma_1(\omega) &=& U^2\,R_{111}(\omega) \  + \
                                 2J^2\,R_{122}(\omega)    \nonumber\\
           &&   +  \  4(U'^2+J^2-U'J)\,R_{212}(\omega)             \\
   {\rm Im}\,\Sigma_{2,3}(\omega) &=& 
                (U^2+2U'^2+3J^2-2U'J)\,R_{222}(\omega)    \nonumber\\
           &&   +  \              J^2\,R_{211}(\omega)    \nonumber\\
           &&   +  \  2(U'^2+J^2-U'J)\,R_{112}(\omega) \ .
\end{eqnarray}
The above expressions demonstrate that even for \,$J=0$\, the self-energy 
of a given band depends on scattering proccesses involving all three 
$t_{2g}$ bands. Nevertheless, $\Sigma_{xy}$ is dominated by interactions 
within the wide $xy$ band or between $xy$ and $xz,yz$. On the other hand,   
$\Sigma_{xz,yz}$ primarily depends on interactions within the narrow
$xz,yz$ bands or between $xz,yz$ and $xy$. These differences are a 
consequence of the layered structure of Sr$_2$RuO$_4$ and give rise to 
anisotropic relaxation shifts.  

For a more accurate description of charge transfer among quasi-particle 
bands, we include self-consistency in the spirit of dynamical mean-field 
theory \cite{georges}. In this scheme, $\Sigma_i$ is a functional of the 
effective bath Green's function ${\cal G}_i^{-1}=G_i^{-1}+\Sigma_i$, where
the local $G_i$ is given by %the Hilbert transform 
\begin{equation} 
     G_i(\omega) = \int_{-\infty}^\infty\!d\omega'
  \ {\rho_i(\omega')\over\omega+\mu-\Sigma_i(\omega)-\omega'}\ .  
\end{equation} 
A typical frequency variation of $\Sigma_i$ is shown in Fig.~2.
Near $E_F$, the imaginary parts vary quadratically with frequency and 
the real parts satisfy \,$\Sigma_{xz,yz}\gg\Sigma_{xy}$, i.e., the energy 
shift of the narrow $xz,yz$ bands is much larger than for the wide $xy$ 
band. Moreover, the difference \,$\Sigma_{xz,yz}-\Sigma_{xy}$\, at $E_F$ 
is much larger than the difference between the Hartree-Fock energies 
\,$\Sigma^{HF}_{xz,yz}-\Sigma^{HF}_{xy}$.

Qualitatively similar results are derived from more refined treatments
of on-site Coulomb correlations using multi-band self-consistent
Quantum Monte Carlo (QMC) methods \cite{georges,rozenberg}.  
The temperature of the simulation was 15~meV with 128 imaginary 
time slices and $\sim300\,000$ Monte Carlo sweeps. 
Fig.~3 shows the quasi-particle density of states 
\,$N_i(\omega)=-{1\over\pi}\,{\rm Im}\,G_i(\omega)$, obtained via
maximum entropy reconstruction \cite{MEM}, together with the bare density 
of states $\rho_i(\omega)$. The van Hove singularities near the edges 
of the $xz,yz$ bands are shifted towards $E_F$, causing a sizeable band 
narrowing. Because of the $\sim2/3$ filling of these bands, this effect 
is not symmetric, giving a stronger relaxation shift of the occupied 
bands than for the unoccupied bands. There is also some band narrowing 
of the $xy$ bands, but since \,$U<W_{xy}$\, this effect is much smaller 
than for the $xz,yz$ bands.

%2

%2
\begin{figure}[t!]
\begin{center}
\vskip -0.5cm 
\epsfig{figure=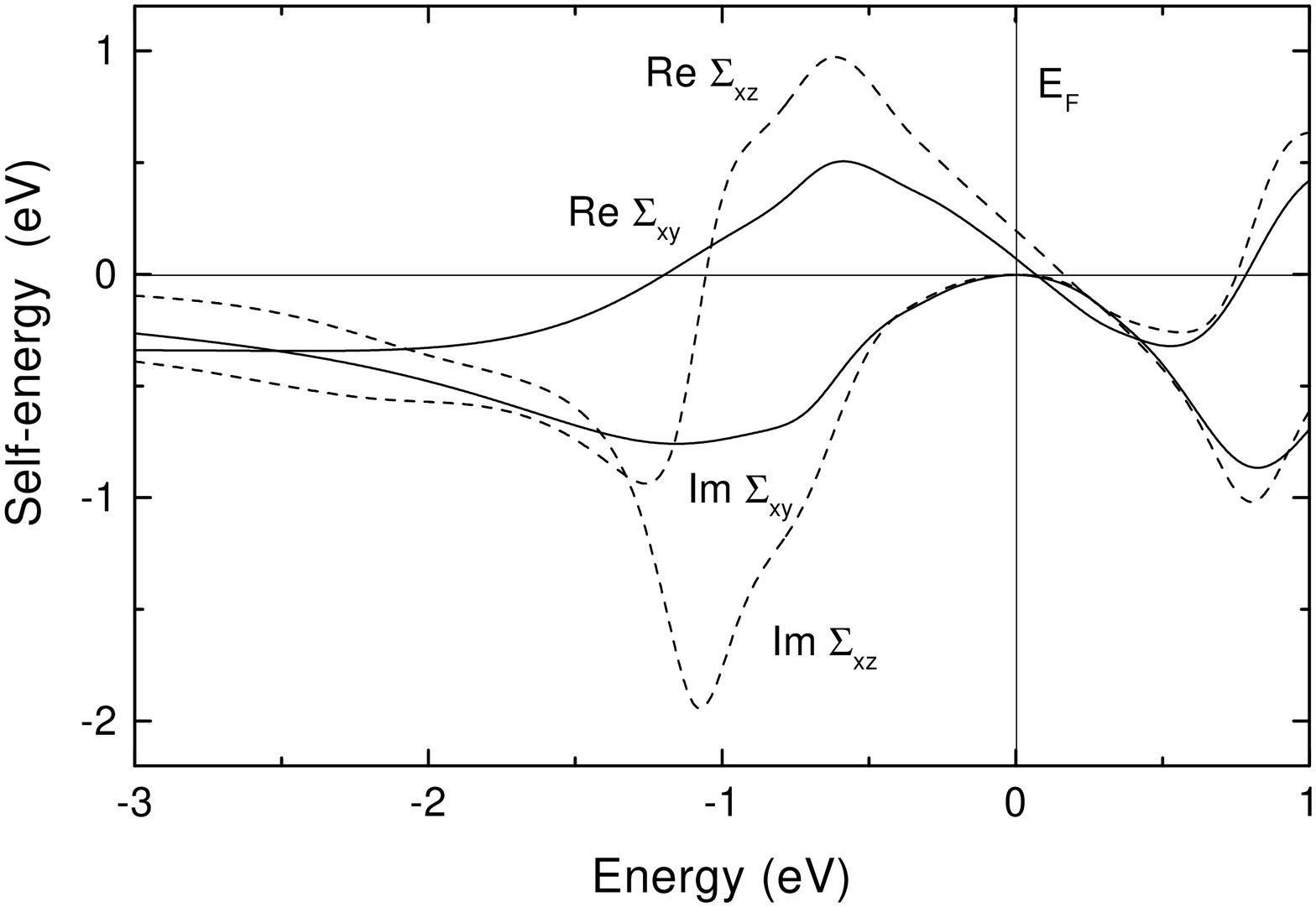,width=8.5cm,height=6.5cm,angle=0}
\vskip -0.4cm
\end{center}
\caption{
Real and imaginary parts of self-consistent second-order self-energy 
for \,$U=1.2$~eV, $J=0.2$~eV. Solid curves: $xy$, dashed curves: $xz$.
}\end{figure}

A crucial point is now that in order to satisfy 
the Luttinger theorem the more pronounced band narrowing of the $xz,yz$ 
bands requires a transfer of spectral weight to the $xy$ bands. Thus, 
the $xy$ van Hove singularity is pushed towards the Fermi level. In the 
example shown in Fig.~3, it lies about 10~meV above $E_F$, compared to 
50~meV in the single-particle spectrum. We emphasize that this result
is a genuine multi-band effect where the filling of a relatively wide 
quasi-particle band is modified by correlations within other narrow 
bands of a different symmetry. Since the values of $U$ and 
$J$ are not well known, and considering the approximate nature of our 
single-particle bands and self-energy calculations, it is not possible 
at present to predict the exact position of the $xy$ singularity. 
It is conceivable, therefore, that this saddle point might lie even 
closer to or below $E_F$. 

As indicated in Fig.~1, the topology of the Fermi surface of Sr$_2$RuO$_4$ 
depends critically on the position of the $xy$ van Hove singularity with 
respect to $E_F$. It is evident therefore that the charge transfer from 
$xz,yz$ to $xy$ due to the creation of the photohole must be taken into 
account when using angle-resolved photoemission to determine the shape 
of the Fermi surface. To compare our results with photoemission spectra, 
we show in Fig.~4(a) the dispersion of the $t_{2g}$ quasi-particle bands 
along $\Gamma M$ and $MX$ derived from the spectral function  
\,$A_i({\bf k,\omega})=-{1\over\pi}\,{\rm Im}\,\big[\omega+\mu-
\varepsilon_i({\bf k})-\Sigma_i(\omega)\big]^{-1}$.
The $xy$ van Hove singularity at $M$ lies 10~meV above $E_F$, so that 
considerable spectral weight appears below $E_F$ in the immediate vicinity 
of $M$. To account for the finite energy resolution, and following the 
experimental procedure for determining the spectral weight near $E_F$ 
\cite{puchkov}, we show in Fig.~4(b) the Fermi surface obtained from 
the partially integrated spectral function 
\,$\bar A_i({\bf k})=\int^\Delta_{-\Delta}\!d\omega\,
A_i({\bf k},\omega+i\Delta)$ with \,$\Delta=25$~meV.
Considering in addition the finite aperture of the detector 
(typically $\pm1^o$, corresponding to $\pm5$\% of $k_\Vert$ near 
$M$ for 25~eV photon energy), it is unavoidable to pick up spectral 
weight from occupied regions near $M$, even when the detector is nominally 
set at $M$. Thus, the near-degeneracy of the $xy$ singularity with $E_F$ 
makes it extremely difficult using angle-resolved photoemission to 
determine the $k$-point at which the $xy$ band crosses the Fermi energy. 
Photoemission data taken with better energy and angle resolution might 
provide a more conclusive answer. 

%3               

%3
\begin{figure}[t!]
\begin{center}
\vskip -0.1cm
\epsfig{figure=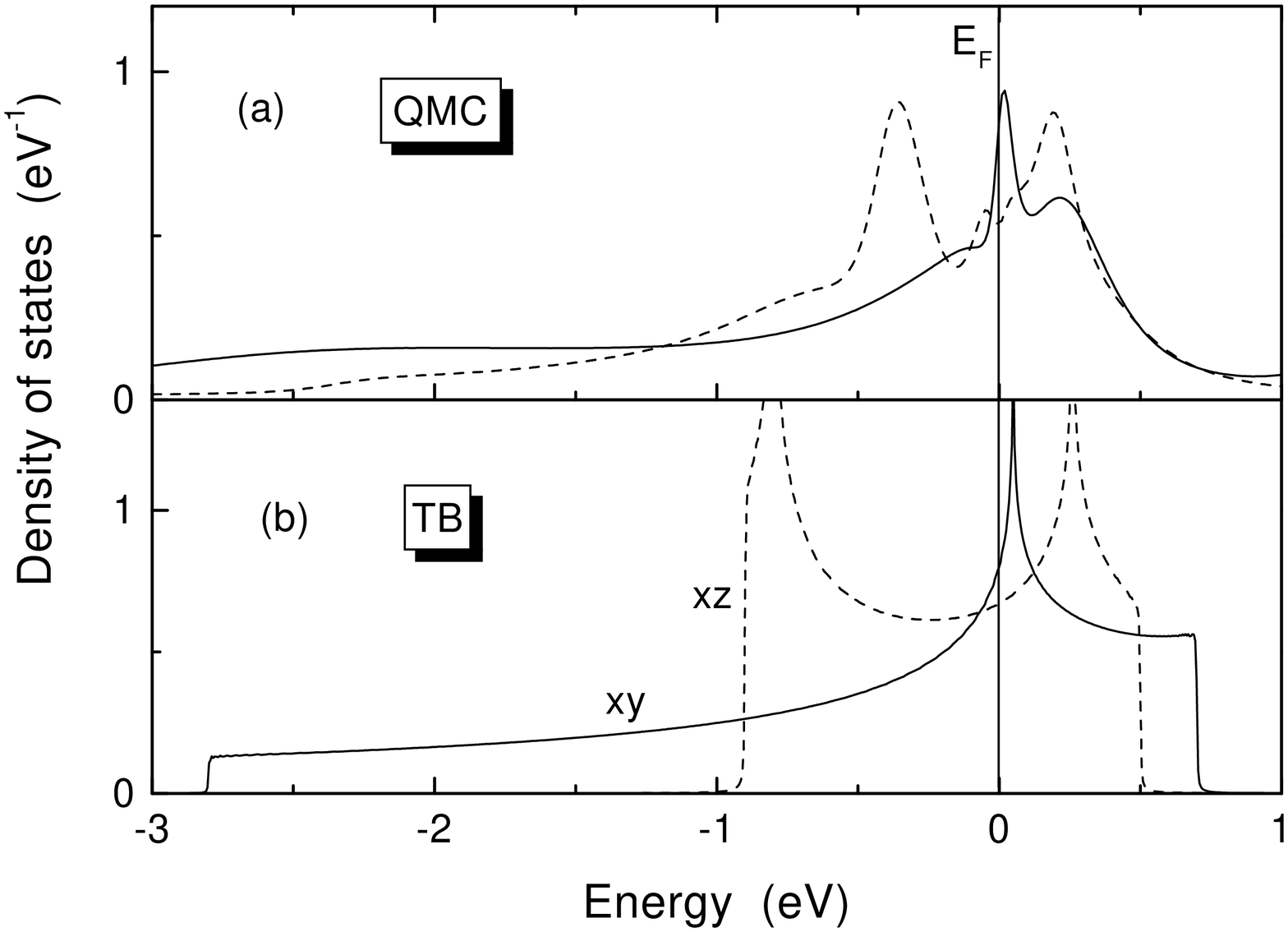,width=8.5cm,height=7.5cm,angle=0}
\vskip -0.4cm
\end{center}

\caption{
(a) Quasi-particle density of states  \,$N_i(\omega)$\, derived from 
self-consistent QMC scheme for \,$U=1.2$~eV, $J=0.2$~eV. 
(b) Single-particle density of states $\rho_i(\omega)$ derived from 
tight binding bands. Solid curves: $xy$, dashed curves: $xz$.
}\end{figure}

Figs.~3 and 4 also show that due to the narrowing of the $xz,yz$ 
bands, the weakly dispersive band is shifted from $-0.8$~eV to about 
$-0.4$~eV, in agreement with photoemission data \cite{yokoya,lu,puchkov}. 
For $k_\Vert$ between $M$ and $X$, this band is observed to 
cross $E_F$ at about $(\pi,0.6\pi)$, in good accord with our calculations.  
In addition, the calculations indicate the existence of a satellite below
the $xz,yz$ bands which might be related to the spectral feature observed 
near 2.5~eV binding energy using resonant photoemission \cite{satellite}. 
The precise location of this satellite is difficult to determine because 
of the uncertainty of $U$ and the approximate nature of our self-energy 
calculations.

%4

%4
\begin{figure}[t!]
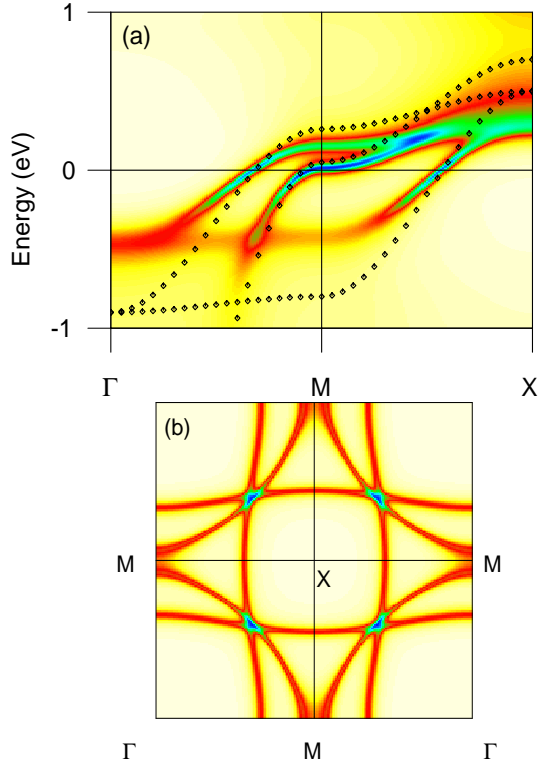

\begin{center}
\vskip -0.5cm 
\epsfig{figure=fig4a.cps,width=6cm,height=8cm,angle=-90}
\vskip -1.1cm
\end{center}
\begin{center}
\epsfig{figure=fig4b.cps,width=6cm,height=6cm,angle=0}
\vskip -0.6cm
\end{center}

\caption{                
(a) Quasi-particle bands along $\Gamma M$ and $MX$ derived from 
self-consistent second-order self-energy. Symbols: tight-binding bands. 
(b) Quasi-particle Fermi surface after accounting for energy 
broadening and resolution (see text).
}\end{figure}

Because of the proximity of the quasi-particle $xy$ van Hove critical point
to the Fermi level, the imaginary part of the self-consistent self-energy 
exhibits a small linear contribution near $E_F$, indicating that the
system may partially behave like a marginal Fermi liquid. In fact, in 
Eq.~(3), it is only the first term $\sim R_{111}(\omega)$ that gives rise 
to a linear term if the singularity coincides with $E_F$. As a result of 
multi-band effects, however, this contribution is rapidly dominated by
stronger quadratic terms involving the narrow $xz,yz$ bands. Thus, we
find the marginality to be rather weak.

We finally discuss the mass renormalization derived from our quasi-particle
bands.  For Coulomb and exchange matrix elements in the range 
\,$U=1.2 - 1.5$~eV, $J=0.2 -0.4$~eV we find \,$m^*/m\approx2.1 - 2.6$, 
in agreement with photoemission estimates \,$m^*/m\approx2.5$ \cite{puchkov}, 
while dHvA measurements \cite{mackenzie} and specific heat data 
\cite{maeno97} suggest a factor of $3-4$.   

In summary, multi-band quasi-particle calculations for Sr$_2$RuO$_4$ show 
that the simultaneous existence of nearly one- and two-dimensional $t_{2g}$ 
bands near $E_F$ leads to a highly anisotropic self-energy of the 
photoemission hole state. Because of Luttinger's theorem, this anisotropy      
gives rise to a charge flow from the narrow $xz,yz$ bands to the wide $xy$
band, thereby shifting the $xy$ van Hove singularity very close to $E_F$.
As a result, in the vicinity of $M$ considerable spectral weight appears
below $E_F$. These results might explain the controversial nature of recent 
photoemission data which have difficulty in determining whether or not 
the $xy$ band at $M$ is occupied.        
 
The calculations were performed on the Cray T3e of the Forschungszentrum 
J\"ulich with grants from the John von Neumann Institute for Computing.

% $^*$e-mails: \\ 
% a.liebsch@fz-juelich.de, \ a.liechtenstein@fz-juelich.de 

\end{multicols}

\begin{references}

\bibitem{maeno}
Y. Maeno, H. Hashimoto, K. Yoshida, S. Nishizaki, T. Fujita,
J.G. Bednorz, and F. Lichtenberg,
Nature {\bf 372}, 532 (1994).

\bibitem{mackenzie}
A.P. Mackenzie, S.R. Julian, A.J. Diver, G.J. MacMullan, M.P. Ray, 
G.G. Lonzarich, Y. Maeno, S. Nishizaki, and T. Fujita, 
Phys. Rev. Lett. {\bf 76}, 3786 (1996);
A.P. Mackenzie, S.R. Julian, G.G. Lonzarich, Y. Maeno, and T. Fujita, 
Phys. Rev. Lett. {\bf 78}, 2271 (1997).

\bibitem{yokoya}
T. Yokoya, A. Chainani, T. Takahashi, H. Katayama-Yoshida, M Kasai, 
and Y. Tokura,
Phys. Rev. Lett. {\bf 76}, 3009 (1996); 
{\it ibid.} {\bf 78}, 2272 (1997);
T. Yokoya, A. Chainani, T. Takahashi, H. Ding, J.C. Campuzano, 
H. Katayama-Yoshida, M Kasai, and Y. Tokura,
Phys. Rev. B {\bf 54}, 13311 (1996).

\bibitem{lu}
D.H. Lu, M. Schmidt, T.R. Cummins, S. Schuppler, F. Lichtenberg,
and J.G. Bednorz, 
Phys. Rev. Lett. {\bf 76}, 4845 (1996).

\bibitem{puchkov}
A.V. Puchkov, Z.X. Shen, T. Kimura, and Y. Tokura,
Phys. Rev. B {\bf 58}, R13322 (1998). 
According to these data, the intensity near $M$ varies strongly 
with photon energy, suggesting that at 22~eV the $xy$ van Hove 
singularity is occupied while above 25~eV it is unoccupied. 

\bibitem{oguchi}
T. Oguchi,
Phys. Rev. B {\bf 51}, 1385 (1995).

\bibitem{mazin}
I.I. Mazin and D.J. Singh,
Phys. Rev. Lett. {\bf 79}, 733 (1997);
D.J. Singh,
Phys. Rev. B {\bf 52}, 1358 (1995).

\bibitem{fang} 
A. Fang and K. Terakura, private communication.

\bibitem{mac98}
A.P. Mackenzie, S. Ikeda,  Y. Maeno, T. Fujita, R. Julian, and 
G.G. Lonzarich, J. Phys. Soc. Japan {\bf 67}, 385 (1998).

\bibitem{satellite}
T. Yokoya, A. Chainani, T. Takahashi, H. Katayama-Yoshida, M Kasai, 
Y. Tokura, N. Shanthi and D.D. Sarma,
Phys. Rev. B {\bf 53}, 8151 (1996).

\bibitem{kanamori}
J. Kanamori, Progr. Theoret. Phys. {\bf 30}, 275 (1963); 
J. Igarashi, P. Unger, K. Hirai, and P. Fulde, 
Phys. Rev. B {\bf 49}, 16181 (1994).

\bibitem{georges}
A. Georges, G. Kotliar, W. Kraut, and M.J. Rozenberg, 
Rev. Mod. Phys. {\bf 68}, 13  (1996).

\bibitem{rozenberg}
M.J. Rozenberg, Phys. Rev. B {\bf 55}, R4855 (1997).

\bibitem{MEM}  
M. Jarrell and J. E. Gubernatis, Phys. Rep. {\bf 269}, 133 (1996).

\bibitem{maeno97}
Y. Maeno {\it et al.}, J. Low Temp. {\bf 105}, 1577 (1997).

\end{references}
\end{document}